\documentstyle[prl,aps,floats,psfig]{revtex}
\draft
\begin{document}

\ifpreprintsty \typeout{}
\else
\twocolumn[\hsize\textwidth\columnwidth\hsize\csname 
@twocolumnfalse\endcsname
\fi

\title{Phases of massive scalar field collapse}
\author{
	Patrick R. Brady,${}^{(1)}$
        Chris M. Chambers,${}^{(2)}$ and 
        S\'ergio M. C. V. Gon\c{c}alves${}^{(3)}$
       }
\address{
         ${}^{(1)}$ Theoretical Astrophysics 130-33,
         	  California Institute of Technology,
		  Pasadena,  CA 91125
        }
\address{
         ${}^{(2)}$ Department of Physics,
		   Montana State University,
        	   Bozeman,
		   MT 59717
        }
\address{
         ${}^{(3)}$ Department of Physics, 
		University of Newcastle upon Tyne, 
		NE1 7RU U.K.
        }

\date{27 May 1997}
\preprint{GRP-475}

\maketitle

\typeout{ABSTRACT}
\begin{abstract}
We study critical behavior in the collapse of massive spherically
symmetric scalar fields. We observe two distinct types of phase
transition at the threshold of black hole formation.  Type II phase
transitions occur when the radial extent $(\lambda)$ of the initial
pulse is less than the Compton wavelength ($\mu^{-1}$) of the scalar
field.  The critical solution is that found by Choptuik in the
collapse of massless scalar fields. Type I phase transitions, where
the black hole formation turns on at finite mass, occur when $\lambda
\mu \gg 1$.  The critical solutions are unstable soliton stars with
masses $\alt 0.6 \mu^{-1}$. Our results in combination with those
obtained for the collapse of a Yang-Mills field~{[M.~W. Choptuik,
T. Chmaj, and P. Bizon, Phys. Rev. Lett. {\bf 77}, 424 (1996)]}
suggest that unstable, confined solutions to the Einstein-matter
equations may be relevant to the critical point of other matter
models.\\
\end{abstract}

\ifpreprintsty \typeout{}
\else
]\narrowtext
\fi

The discovery of critical point behavior in gravitational collapse has
highlighted the role played by non-linear dynamics at the threshold of
black hole formation, and has opened up a fascinating area of research
in General Relativity.

Choptuik~\cite{Choptuik_M:1993} performed the first definitive
numerical study of critical behavior in the collapse of
spherically-symmetric distributions of massless scalar field. His
results indicated that one parameter families of interpolating
solutions $S[p]$ generically have a critical value $p=p^{*}$ such that
(i) $S[p < p^{*}]$ are solutions in which the scalar field disperses
to infinity, and (ii) $S [ p > p^{*} ]$ are solutions in which the
field collapses to form a black hole.  In slightly super-critical
evolutions, Choptuik found that the black-hole mass has a simple
power-law form 
\begin{equation} 
	M_{BH} \simeq K |p-p^*|^\gamma \; , \label{eq:0}
\end{equation} 
where the critical exponent is $\gamma \simeq 0.37$ and $K$ is a
family dependent constant.  For near critical evolutions the field
asymptotically approaches a discretely self-similar form, with an
echoing period $\triangle \simeq 3.44$ which is the same for all
families, before either dispersing to infinity or forming a black
hole.  Based on these observations, Choptuik conjectured that a unique
solution to the Einstein-scalar field equations acts as the
intermediate attractor for near critical evolutions.  Gundlach has
directly constructed this critical solution, and has computed the
echoing period and the critical exponent to be $\triangle = 3.4453\pm
0.0005$ and $\gamma = 0.374
\pm0.001$~\cite{Gundlach_C:1997,Gundlach_C:1995}, thus confirming the
numerical estimates.  It has also been argued that this picture is
stable against the introduction of a small scalar field
mass~\cite{Gundlach_C:1995,Hirschmann_E:1995,Choptuik_M:1994}.

Motivated by the results of Choptuik, critical point behavior has also
been studied in other models of gravitational
collapse~\cite{Evans_C:1994,Abrahams_A:1993,Hod_S:1997a,Liebling_S:1996}.
Dynamical self-similarity (either discrete or continuous) in near
critical evolutions, and a scaling relation for black-hole mass, as in
Eq.~(\ref{eq:0}), are common features of these models, although the
numerical value of the critical exponent $\gamma$ is model dependent.
It is now well established that the power law form for the black-hole
mass derives from the existence of a single unstable mode of the
critical solution in each
case~\cite{Koike_T:1995,Maison_D:1996,Gundlach_C:1997}.  Of the
examples considered to date, the evolution of the Yang-Mills field is
exceptional.  Choptuik {\it et al}~\cite{Choptuik_M:1996} have studied
this model, finding two distinct types of phase transition depending
on the initial field configurations they considered.  In what they
refer to as {\em Type I} transitions, the black hole formation turns
on at finite mass and the critical solution is the Bartnik-McKinnon
solution~\cite{Bartnik_R:1988}---a regular, static, but unstable,
solution to the spherically symmetric Einstein-Yang-Mills equations.
In {\em Type II} transitions black-hole formation turns on at
infinitesimal mass and the critical behavior is qualitatively similar
to that found by Choptuik for massless scalar fields, except that the
scaling exponent is $\gamma
\simeq 0.20$ and the echoing period is $\triangle \simeq 0.74$.  An
independent confirmation of these results has been provided by
Gundlach~\cite{Gundlach_C:1997a}.    

Here we report on a detailed study of critical phenomenon in the
collapse of spherically symmetric configurations of a massive scalar
field.  The introduction of a mass $\mu$ destroys the scale invariance
of the Einstein-scalar field equations.  Moreover, the massive
scalar-field equations admit soliton-like solutions as discussed by
Seidel and Suen~\cite{Seidel_E:1991}.  These observations suggest that
the qualitative picture of critical point behavior could differ from
the massless limit, and might be similar to that found by Choptuik
{\it et al.}~\cite{Choptuik_M:1996} in their study of
Einstein-Yang-Mills collapse\footnote{In this case, the SU(2)
Yang-Mills charge breaks the scale invariance of the field
equations.}.  Our results show that both Type I and Type II phase
transitions occur in the collapse of massive scalar fields.
Furthermore, we advance a simple criterion to determine which type of
phase transition will be observed for a given initial data set. If the
radial extent $\lambda$ of the initial pulse is greater than the
Compton wavelength $\mu^{-1}$ of the scalar field then Type~I phase
transitions will be observed.  Type~II transitions develop from
initial data with $\lambda\mu \alt 1$.  This criterion provides an
intuitive, physical explanation of the observed phenomenology, and
clarifies the role played by intrinsic scales in critical collapse.

We write the general, spherically symmetric line-element in terms of a
retarded time $u$ and a radial coordinate $r$, which measures proper
area of the 2-spheres, as
\begin{equation}
	ds^2 = -g \overline{g} du^2 - 2gdudr + r^2 d\Omega^2 \; ,
	\label{eq:1}
\end{equation}
where $g$ and $\overline{g}$ are functions of both $r$ and $u$, and
$d\Omega^2$ is the line-element on the unit $2$-sphere.  We choose to
normalize $u$ to be proper time at the origin, thus fixing $g(0,u)=1$.
Imposing regularity of the spacetime at the origin requires
$\overline{g}(0,u)=1$.  The evolution of the massive scalar field is governed
by the wave equation $\Box \phi - \mu^2 \phi = 0$.  It is convenient
to introduce an auxiliary field $h$ \cite{Christodoulou_D:1984}, related
to $\phi$ by
\begin{equation}
	\phi=\overline{h}=\frac{1}{r} \int_0^r dr' h(r',u) \; ,
\end{equation}
in terms of which the wave equation can easily be written in a first
order form.  Thus, the coupled Einstein-scalar field equations are
\begin{eqnarray}
	(\ln g)_{,r} &=& r^{-1} (h-\overline{h})^2 \; ,\\
	(r\overline{g})_{,r} &=& g(1-\mu^2 r^2 \overline{h}^2) \; , \\
	(r\overline{h})_{,r} &=& h \; , \\
	h_{,u} - \frac{\overline{g}}{2}h_{,r} &=& \frac{1}{2r}
			(h-\overline{h})[g(1 - \mu^2 r^2 \overline{h}^2)
			-\overline{g}] \nonumber\\
			&& \ \ - r g \mu^2 \overline{h} /2 \; .
\end{eqnarray}
The dynamics is completely encompassed in the last equation, which is
the wave equation written in terms of the new fields $h$ and
$\overline{h}$.

The characteristic initial value problem requires only the field
$\phi$ to be supplied on some initial outgoing null cone, which we
will take, without loss of generality, to be at $u=0$. We have
considered the evolution of three different initial data sets as shown
in Table~\ref{table:1}.

\ifpreprintsty 
\typeout{Putting table at the end}
\else
\begin{table}[h]
\caption{
The three initial data sets considered in our evolutions.  Only a single
parameter is varied when looking for a critical point.  The types of
phase transitions which may occur are indicated under Type.  }
\vskip\baselineskip
\begin{tabular}{cccc}
&$\phi(u=0,r)$ & Parameters & Type \\\hline\\
(i) & $\phi_0 r^2 \exp\left[-{(r-r_0)^2}/{\sigma^2}\right]$& $\sigma$,
$\phi_0$& I, II \\
(ii) &$\phi_0 \left\{1-\tanh[(r-r_0)/\sigma]\right\}$ & 
	$\sigma$, $\phi_0$ & I, II \\
(iii) &$\phi_0 r (r+r_0)^{-\sigma}/ (1+e^r)$ & $\sigma$, $\phi_0$ & I,
II \\
\\
\end{tabular}
\label{table:1}
\end{table}
\fi

The numerical algorithm used to integrate these equations is
documented~\cite{Goldwirth_D:1987,Garfinkle_D:1994} elsewhere.  We
have followed the scheme as outlined by
Garfinkle~\cite{Garfinkle_D:1994}.  The accuracy of the code has been
tested previously, where it was used to study radiative tails of a
massless scalar field propagating in asymptotically de~Sitter
spacetimes~\cite{Brady_P:1997}, and was found to be locally second
order accurate.

It is reasonable to expect that the picture of critical behavior
offered by Choptuik~\cite{Choptuik_M:1993} is robust against the
introduction of a small scalar-field
mass~\cite{Gundlach_C:1995,Hirschmann_E:1995,Choptuik_M:1994}.  More
precisely, the evolution of an initial distribution of scalar field
will differ from the massless evolution only if the characteristic
length-scale $\mu^{-1}$, set by the scalar field mass, is smaller than
the radial extent $\lambda$ of the region in which the field is
non-zero\footnote[1]{For generic initial data it is not possible to
define the radial extent, however, for the initial data sets (i) and
(ii) in Table~\protect\ref{table:1}, we set $\lambda = 2
\sigma$.}.  This expectation is supported by our numerical
integrations.  We generally observe Type II phase transitions when
$\lambda\mu \ll 1$.  Furthermore, we find that the scaling relation
for the black hole mass is in agreement with the massless limit,
having $\gamma \approx 0.378$.

The new feature, arising due to the presence of the mass $\mu$, is the
existence of Type~I phase transitions---phase transitions in which
black hole formation turns on at finite mass.  The critical solutions
are soliton stars on the unstable branch of the mass versus radius
curve discussed by Seidel and Suen~\cite{Seidel_E:1991}.  The mass gap
at the threshold of black hole formation lies in the range $0.35 \alt
\mu\, M_{BH} \alt 0.59 $, the upper limit being set by the maximum
mass that a soliton star can have.

For the three initial data sets we examined (see Table~\ref{table:1}),
we have found both Type I and Type II behavior, along with evidence
that both critical solutions play a role when $\lambda\mu \approx 1$.
In contrast to the Einstein-Yang-Mills system~\cite{Choptuik_M:1996}
our results suggest that the shape of the initial data does not
determine the critical point behavior.

Physically the existence of the different regimes can be understood if
we recall the two known limits for scalar field collapse. In the
massless regime $\lambda \mu \ll 1$ an outward pressure is required
for the field to bounce back to infinity, whereas in the adiabatic
regime $\lambda\mu \gg 1$ the collapse is
pressureless~\cite{Goncalves_S:1997}.  Hence, by continuity in the
space of solutions, it seems likely that there can exist
configurations, characterized by $\lambda\mu=C\sim {\cal O}(1)$, such
that the field neither disperses to infinity nor collapses to a
singularity.

Figure~\ref{fig:2} shows the field $\phi$ at the origin $r=0$ during
the critical phase of a Type~I evolution for the Gaussian initial data
in Table~\ref{table:1} with $r_0=5.0$, $\sigma=1.2$ and $\phi_0 =
6.4746$.  The solution behaves like a soliton star with an effective
mass $\sim 0.52\mu^{-1} $ confined within a radius $\sim 4 \mu^{-1} $
for an amount of time $u\sim100$.  The angular frequency of its
fundamental mode is $\omega_{0} \approx 1.8 \mu$, the next mode is
also apparent at $3\,\omega_{0}$.  Superimposed on these oscillations
is an amplitude modulation with period $\sim 15.7 \mu^{-1}$ which is
reflected in the sideband structure of the Fourier amplitude.

Sub-critical evolutions generally settle down to a scalar field
configuration dominated by a single oscillatory mode with angular
frequency $\omega \approx 1.05$ as shown in Fig~\ref{fig:3} for $\mu =
1.0$.  Further exploration indicates that the fundamental oscillations
have a period given approximately by $2\pi\mu^{-1}$.  Unfortunately,
the characteristic evolution scheme makes it
difficult to follow the evolutions beyond $u\sim 400$.  Nevertheless,
the trend suggests that the amplitude of the scalar field is
decreasing slowly---indeed this is precisely the regime in which the
stationary phase approximation is valid~\cite{Goncalves_S:1997}.

When the critical solution corresponds to the marginally stable
soliton star with effective mass $\sim 0.6 \mu^{-1}$, we have found
evidence of further phenomenology.  In particular, the solutions may
closely approach the solitonic configuration, begin to disperse, but
recollapse to form black holes.  This behavior merits further
investigation,  however our numerical scheme is not well suited for
this purpose.

As evidence of the observed mass gap, we present in Fig.~\ref{fig:4}
the spectrum of black hole masses near to criticality for the initial
data set (iii) of Table~\ref{table:1}.  The black-hole mass at
threshold is $M_{BH} \approx 0.51\mu^{-1}$.  This mass spectrum is
most interesting for what it does not do, rather than what it does.
We determine that a solution contains a black hole if either of the
metric functions $g$ or $\overline{g}$ exceeds some pre-specified
tolerance $G_{\rm max}$ anywhere on a slice of constant $u$.  Suppose
this occurs at $u=u_{BH}[\phi_0]$, where $\phi_0$ is the parameter
being varied in the initial data.  The mass of the black hole is then
$M_{BH} = \frac{1}{2} r_{BH}$ where $r_{BH}$ is the location of the {\em
global} minimum of $f(r;\phi_0) \equiv
\overline{g}(r,u_{BH}[\phi_0]) / g(r,u_{BH}[\phi_0])$ on the 
slice $u=u_{BH}$.  Notice that $M_{BH}$ need not depend continuously
on $\phi_0$ if $f(r;\phi_0)$ has more than one local minimum.  Indeed,
as $\phi_0$ is varied in our simulations we sometimes observe
discontinuities in the black-hole mass when the tolerance $G_{\rm
max}$ is not large enough; this is shown in the inset of
Fig.~\ref{fig:4}. A careful inspection, varying both the numerical
resolution and the tolerance, suggests that the discontinuities are
{\em not} a real effect in the black-hole mass spectrum.
(Discontinuities in the mass spectrum are also alluded to
in~\cite{Choptuik_M:1996}; it would be interesting to check if they
arise for similar reasons.)  In contrast, the oscillation imposed on
the mass spectrum in Fig.~\ref{fig:4} is not an artifact of the
numerics but is similar to the fine structure found by Hod and
Piran~\cite{Hod_S:1997} in the Choptuik results.

To better understand the selection effect between Type I and Type II
phase transitions we have constructed families of interpolating
solutions $S_\lambda[\phi_0]$ for several values of $\lambda$.
Generally, we find Type I transitions occur when the Bondi mass
$M_{\rm Bondi}$ of the initial field profile is greater than $\sim 0.4
\mu^{-1}$ and its radial extent is larger than the Compton wavelength
of the field, i.e. $\lambda \mu \agt 1$.  Figure~\ref{fig:1} shows the
Bondi mass of the initial data at the critical point
$\phi_0=\phi_0^*$, and the resulting black-hole mass, for the initial data
sets (i) and (ii) in Table~\ref{table:1}.

In conclusion, we find that the presence of a length scale changes the
nature of critical phenomena in gravitational collapse of a scalar
field. It introduces new phenomenology which is similar to that
discussed by Choptuik et al~\cite{Choptuik_M:1996}.  Moreover, it is
tempting to speculate that unstable, confined solutions will act as
critical solutions in other matter models.  We therefore expect that
both Type~I and Type~II phase transitions should occur in the
gravitational collapse of perfect fluids (with equations of state
which allow stationary configurations), {\em and} in the collapse of
charged massive scalar fields~\cite{Seidel_E:1990}.

We are grateful to Thomas Brueckner for useful conversations.  PRB is
supported in part by NSF grant AST-9417371, and by a PMA Division
Fellowship at Caltech.  CMC is supported by the Royal Commission For
The Exhibition Of 1851 and gratefully acknowledges their financial
support, and SG is supported by the Programa PRAXIS XXI of the
J.N.I.C.T. of Portugal.


\ifpreprintsty
\clearpage
\begin{table}[h]
\caption{
The three initial data sets considered in our evolutions.  Only a single
parameter is varied when looking for a critical point.  The types of
phase transitions which may occur are indicated under Type.  }
\vskip\baselineskip
\begin{tabular}{cccc}
&$\phi(u=0,r)$ & Parameters & Type \\\hline\\
(i) & $\phi_0 r^2 \exp\left[-{(r-r_0)^2}/{\sigma^2}\right]$& $\sigma$,
$\phi_0$& I, II \\
(ii) &$\phi_0 \left\{1-\tanh[(r-r_0)/\sigma]\right\}$ & 
	$\sigma$, $\phi_0$ & I, II \\
(iii) &$\phi_0 r (r+r_0)^{-\sigma}/ (1+e^r)$ & $\sigma$, $\phi_0$ & I,
II \\
\\
\end{tabular}
\label{table:1}
\end{table}
\fi

\clearpage 

\newdimen\figwidth
\setlength\figwidth{7cm}
\ifpreprintsty
	\setlength\figwidth\textwidth
\fi

\begin{figure}
\psfig{file=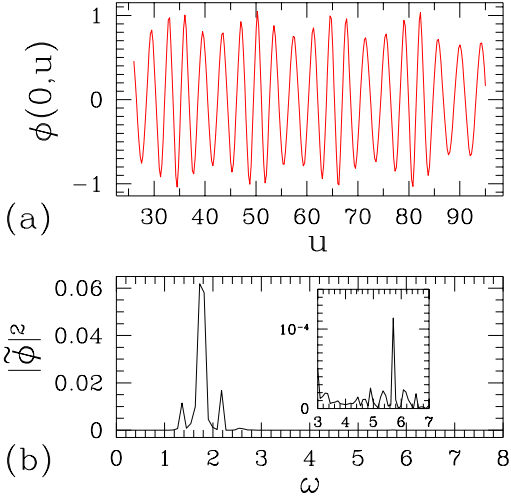,width=\figwidth,bbllx=18pt,bblly=144pt,bburx=592pt,bbury=718pt}
\caption{(a) The time evolution of $\phi$ as a function of retarded
time $u$ at the origin for a near critical evolution with Gaussian
initial data (See Table~\protect\ref{table:1}).  It clearly
exhibits an underlying periodic solution with a superimposed amplitude
modulation. (b) The squared amplitude of the discrete Fourier
transform of $\phi(0,u)$.  The fundamental oscillation has an
angular frequency $\omega_{0} \approx 1.8 \mu$, and, in agreement with
Seidel and Suen\protect\cite{Seidel_E:1991}, the next important
feature is at three times this frequency $\omega\sim 5.65 \mu$.  The
sidebands determine the period of the amplitude modulation to be
$\approx 15.7 \mu^{-1}$.}
\label{fig:2}
\end{figure}

\begin{figure}
\psfig{file=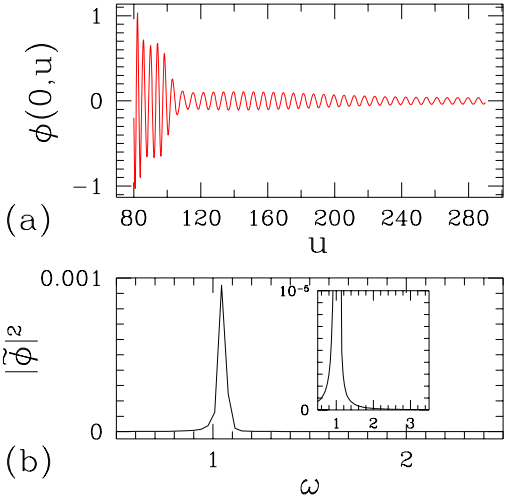,width=\figwidth,bbllx=18pt,bblly=144pt,bburx=592pt,bbury=718pt}
\caption{(a) The time evolution of $\phi$ as a function of retarded
time $u$ at the origin for the same evolution as in
Fig~\protect\ref{fig:2}, but after the nearly critical phase which
ends at $u\approx 100$.  The solution is essentially periodic, with a
low frequency amplitude modulation. (b) The squared amplitude of the
discrete Fourier transform of the field for $u\ge 100$.  The
fundamental oscillation has an angular frequency $\omega_{0} \sim
1.05\mu$, which suggests that it is determined by the scalar field
mass (which is set to unity).  The inset demonstrates that other
harmonics are not strongly excited in this solution.  For other
initial data sets, we find that the amplitude of the oscillations
tends to decay slowly, and the solutions are dispersing.}
\label{fig:3}
\end{figure}

\begin{figure}
\psfig{file=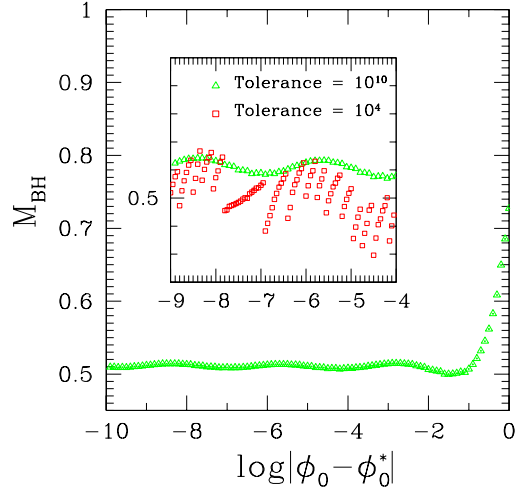,width=\figwidth,bbllx=18pt,bblly=130pt,bburx=592pt,bbury=718pt}
\caption{The black-hole mass $M_{BH}$ as a function of 
$\log | \phi_0 - \phi^{*}_{0} | $ for supercritical evolutions with
$\mu = 1.0$. The results displayed are for the initial data set (iii)
in Table~\protect\ref{table:1} with $r_0=2.0$ and $\sigma=10.0$.  The
critical point was determined to be $\phi_0^* = 3.87245233459$ with an
initial radial discretization $\Delta r = 0.05$. The black hole
tolerance parameter (see text) was set at $10^{10}$. The inset shows
results obtained from the same evolution, but with a lower black hole
tolerance level of $10^{4}$. For low values of the tolerance the mass
spectrum exhibits {\em spurious} discontinuities.}
\label{fig:4}
\end{figure}

\begin{figure}
\psfig{file=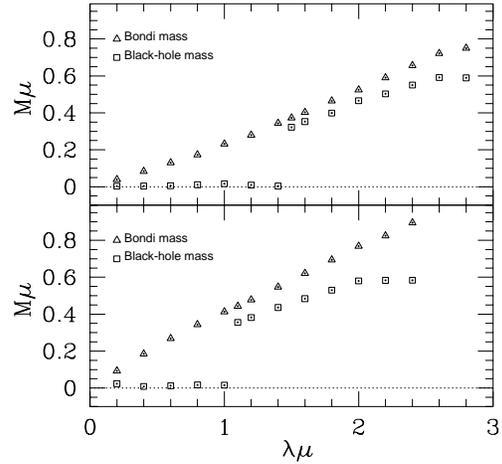,width=\figwidth,bbllx=18pt,bblly=144pt,bburx=592pt,bbury=718pt}
\caption{The Bondi mass of the initial scalar field profile,  in
practice $m(r,0)= r(1 - \overline{g}/g)/2$ evaluated at the outer edge of the
grid, and the measured black-hole mass at the critical point versus
the radial extent $\lambda$ of the initial profile for the
data sets (i) and (ii) in Table~\protect\ref{table:1}.  Type I
transitions are evident for $\lambda \mu \gg 1$, and Type II
transitions when $\lambda \mu \ll 1$.  The interface between
Type~I and Type~II behavior is clearly visible when $M_{\rm Bondi}
\simeq 0.4 \mu^{-1}$,  and $\lambda \mu \sim 1$.   }
\label{fig:1}
\end{figure}

\end{document}